\documentclass[pra,showpacs,priprent,twocolumn,superscriptaddress, floatfix]{revtex4-1}
\usepackage{mathrsfs}
\usepackage{amsfonts}
\usepackage{amsmath}
\usepackage{txfonts}
\usepackage{amssymb} 
\usepackage{graphicx}
\usepackage{bm}
\usepackage{color}
\usepackage[normalem]{ulem}
\usepackage[retainorgcmds]{IEEEtrantools}

\usepackage[colorlinks=true,linkcolor=blue,urlcolor=blue,citecolor=blue,pdfusetitle]{hyperref}

\usepackage{tikz}
\usepackage[mode=buildnew]{standalone}
\usetikzlibrary{decorations.pathmorphing}
\usetikzlibrary{decorations.markings}
\definecolor{LimeGreen}{HTML}{32CD32}

\newcommand{\ket}[1]{|#1\rangle}
\newcommand{\bra}[1]{\langle #1|}

\DeclareMathOperator{\sech}{sech}

\begin{document}

\title{Time optimal holonomic quantum  computation}

\author{Gabriel O. Alves}
\email{alves.go.co@gmail.com}

\affiliation{Instituto de F\'isica da Universidade de S\~ao Paulo,  05314-970 S\~ao Paulo, Brazil.}
\affiliation{Department of Physics and Astronomy, Uppsala University,
Box 516, Se-751 20 Uppsala, Sweden}

\author{Erik Sj\"oqvist}
\email{erik.sjoqvist@physics.uu.se}

\affiliation{Department of Physics and Astronomy, Uppsala University,
Box 516, Se-751 20 Uppsala, Sweden}
\date{\today}

\begin{abstract}
A three-level system can be used in a $\Lambda$-type configuration in order to construct 
a universal set of quantum gates through the use of non-Abelian non-adiabatic 
geometrical phases. Such construction allows for high-speed operation times which diminish 
the effects of decoherence. This might be, however, accompanied by a breakdown of the 
validity of the rotating wave approximation (RWA) due to the comparable time scale between 
counter-rotating terms and the pulse length, which greatly affects the dynamics. Here, 
we investigate the trade-off between dissipative effects and the RWA validity, obtaining the 
optimal regime for the operation of the holonomic quantum gates.
\end{abstract}

\maketitle

\date{\today}

\section{Introduction}

The implementation of a concrete quantum computing platform poses several challenges. 
A suitable platform should exhibit features such as scalability \cite{DiVincenzo2000}, long 
decoherence time \cite{Joos1985, Unruh1995}, and universality \cite{Barenco1995}. In this 
spirit, holonomic quantum computation (HQC) \cite{Zanardi1999,Pachos1999,Duan2001} arose 
as a promising approach for quantum computing. HQC was initially based on the use of adiabatic 
geometric phases, which are robust against dynamical details and fluctuations in the evolution 
\cite{Solinas2012, Viotti2021}. 

 The physical implications of geometrical phases came into the limelight after Berry's seminal work \cite{Berry1984} on adiabatic systems. Since then, several generalizations have 
been made. Of particular importance is the generalization to non-adiabatic non-Abelian 
geometrical phases \cite{Anandan1988}, which have been used in non-adiabatic holonomic 
quantum computing (NHQC), originally proposed in Ref.~\cite{Sjoqvist2012} for a 
$\Lambda$-type system, and which generalizes quantum computation based on non-adiabatic Abelian geometric phases  \cite{Wang2001,Zhu2002} to the non-Abelian case. Due to the long operation time required by adiabatic implementations, 
quantum gates become more susceptible to open quantum system phenomena. Therefore, 
non-adiabatic constructions are often regarded as effective strategies to mitigate this effect 
\cite{Sjoqvist2012,Johansson2012,Shen2021}. Among experimental realizations, implementations 
using superconducting qubits \cite{AbdumalikovJr2013,Danilin2018} and nitrogen-vacancy 
centers in diamond  \cite{Arroyo-Camejo2014,Zu2014} can be found as examples. Moreover, 
single-loop implementations, which speed up the protocol even further, have been investigated 
\cite{Xu2015,Sjoqvist2016, Herterich2016, Zhou2017, Xu2018a,Xu2018b,Chen2020,Han2020}.

Ideally, the $\Lambda$-type system usually operates in the regime of the rotating wave 
approximation (RWA). Whenever the counter-rotating frequencies associated with the bare 
Hamiltonian are large enough in comparison with the typical time scale of the system 
dynamics, the counter-rotating contributions to the Hamiltonian can be averaged out. This, 
in practice, introduces a limitation into how fast these gates can operate \cite{Spiegelberg2013}. 
While arbitrarily fast gates dispel the effects of decoherence, they leave the regime of validity 
of the RWA: the operation time of the gate becomes comparable with the oscillation frequency 
of the counter-rotating terms. These two competing effects introduce a trade-off between 
dissipative effects in the gate and the RWA accuracy. One may note that a somewhat 
analogous trade-off effect occurs in the adiabatic version of HQC, but with the breakdown 
of RWA replaced by non-ideal effects associated with the finite run time of the gates 
\cite{Florio2006a,Florio2006b,Lupo2007}. 

Our objective in this work is to investigate the trade-off between dissipative effects and 
breakdown of the RWA. We analyze how different parameters affect the performance of the 
gates and we also determine the optimal regime of operation for single and two-qubit 
non-adiabatic holonomic quantum gates in the $\Lambda$-type configuration. We find 
that the counter-rotating terms introduce a coupling between the dark and the excited state, which does not occur in the RWA regime. Moreover, we also investigate the effect of 
heterogeneous counter-rotating frequencies. We observe that different frequencies may, 
very slightly, improve the fidelity for certain gates. 

The paper is organized as follows. In Sec.~\ref{sec:holonomic_setting}, we briefly review 
the holonomic setting for one and two-qubit gates. In Sec.~\ref{sec:nonRWA}, we extend 
the discussion beyond the RWA regime and we discuss our results. Concluding remarks 
can be found in Sec.~\ref{sec:conclusions}.

\section{Holonomic setting}\label{sec:holonomic_setting}

In the $\Lambda$-type system, two states $\ket{0}$ and $\ket{1}$, which encode the qubit 
space, are coupled to an auxiliary excited state $\ket{e}$, but remain uncoupled between 
themselves. The system acquires a $\Lambda$-like structure, as depicted in 
Fig.~\ref{fig:setup_dissipative}(a). We can regard the states $\ket{0}$ and $\ket{1}$ as 
stable ground states. Meanwhile, $\ket{e}$ is typically an unstable state that undergoes 
dissipation, decaying to an auxiliary ground state $\ket{g}$. Transitions between the levels 
are induced by a pair of laser pulses, which can be controlled over time.

\subsection{One-qubit gates} 

The starting point to model the unitary dynamics of the single-qubit gates is the Hamiltonian 
\cite{Spiegelberg2013}:
\begin{eqnarray}
H(t) = H_0 + \boldsymbol{\mu} \cdot {\bf E}(t),
\end{eqnarray}
where $H_0 = -f_{0e} \ket{0}\bra{0} - f_{1e} \ket{1}\bra{1}$ is the bare Hamiltonian and
\begin{eqnarray}
{\bf E}(t) = g_0(t)\cos(\nu_0 t) \boldsymbol{\epsilon}_0+ g_1(t) \cos(\nu_1 t) \boldsymbol{\epsilon}_1,    
\end{eqnarray}
is the applied oscillating electric pulse. Here, $g_j(t)$ and $\nu_j$ (with $j=0$ and $1$) are the pulse 
envelope and the oscillation frequency, respectively. Additionally, $\boldsymbol{\mu}$ is the electric 
dipole operator and $\boldsymbol{\epsilon}_j$ is the polarization. We move to the interaction 
picture Hamiltonian $H_I(t) = e^{-i H_0 t}H(t)e^{i H_0 t}$, tuning the frequencies $\nu_j$ so 
they get resonant with the bare transition frequencies $f_{je}$, i.e., $\nu_j = f_{je}$. By doing 
so, one finds the Hamiltonian that describes the $\Lambda$-type system in the interaction 
picture:
\begin{eqnarray}\label{eq:Lambda_hamiltonian}
H_I(t) & = &
\Omega_0(t) (1 + e^{-2if_{0e} t})  \ket{e}\bra{0} \nonumber \\ 
& & +  \Omega_1(t) (1 + e^{-2if_{1e} t})  \ket{e}\bra{1} + {\rm H.c.},  
\end{eqnarray}
where $\Omega_j (t) = \bra{e} \boldsymbol{\mu}  \cdot \boldsymbol{\epsilon} \ket{j} g_j(t)/2$ are 
Rabi frequencies (we put $\hbar = 1$ from now on). Henceforth, we are interested in how to 
handle the counter-rotating terms $e^{-2if_{je} t}$ and how they affect the performance of this 
protocol.

We start by reviewing the ideal case, where the RWA is valid, assuming that $f_{je}$ is 
large. These rapidly oscillating terms average out to zero and the Hamiltonian in 
Eq.~\eqref{eq:Lambda_hamiltonian} becomes:
\begin{eqnarray}\label{eq:Lambda_hamiltonian_rwa}
H_I^{\rm RWA}(t) 
=
\Omega (t) (\omega_0 \ket{e}\bra{0} + \omega_1 \ket{e}\bra{1} )  + {\rm H.c.}  
\end{eqnarray}
Here, we have assumed that both laser pulses are applied simultaneously and have identical 
shape so that we may rewrite the frequencies as $\Omega_j(t) = \Omega(t) \omega_j$, with $\omega_j$  being time-independent and satisfying 
$|\omega_0|^2 + |\omega_1|^2 = 1$. The parameter $\Omega(t)$ can be regarded as an 
overall pulse envelope, while $\omega_0$ and $\omega_1$ refer to the relative (complex) 
weight between the two transition amplitudes.  It is  elucidating to rewrite the Hamiltonian 
above in terms of the bright and the dark states, $\ket{b} = \omega_0^{\ast}\ket{0} + 
\omega_1^{\ast} \ket{1}$ and $\ket{d} = -\omega_1 \ket{0} + \omega_0 \ket{1}$, respectively. 
The Hamiltonian in Eq.~\eqref{eq:Lambda_hamiltonian_rwa} can thereby be re-expressed as
\begin{eqnarray}\label{eq:Lambda_hamiltonian_rwa_bd}
H_{I}^{\rm RWA}(t) = \Omega (t) \ket{e}\bra{b} + {\rm H.c.}
\end{eqnarray}
This means that the dark state is decoupled from the evolution and the system simply performs 
Rabi oscillations between the bright and the excited states with frequency $\Omega(t)$.

\begin{figure}
    \centering
    \includegraphics[width=.45\textwidth]{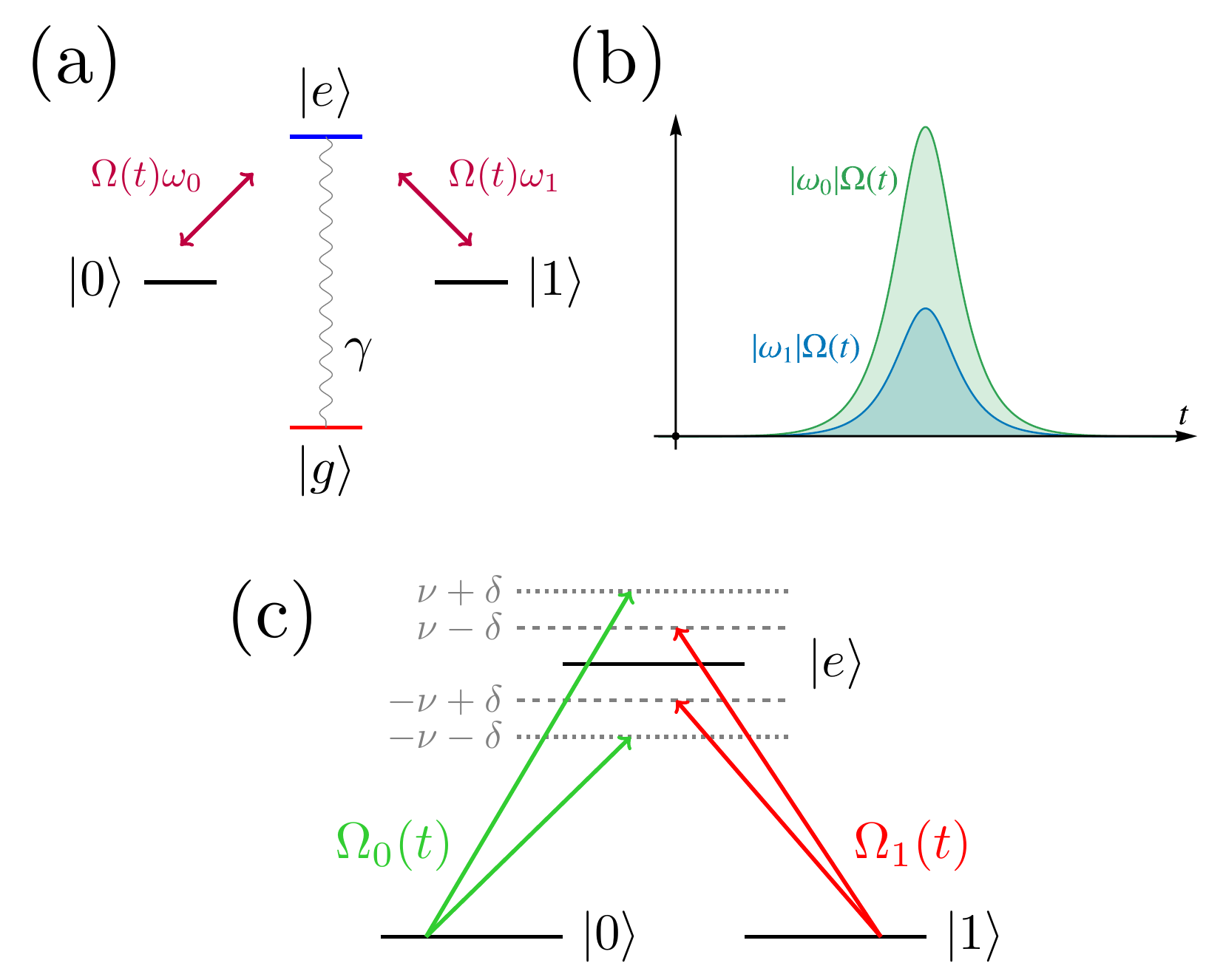}
    \caption{(a) Basic setup for the $\Lambda$-type system. The excited state decays to an 
    additional low-lying level $\ket{g}$ with a rate $\gamma$. There is no coupling or decay 
    with the computational subspace itself. This means that dissipation effects only occur 
    while the excited state is populated. (b) Hyperbolic secant pulses as defined in Eq.~\eqref{eq:pulse_shape}. (c) Configuration and 
    detunings in the two-qubit gate setup.}
    \label{fig:setup_dissipative}
\end{figure}

Let us now consider the qubit subspace $M(0) = \text{Span} \{\ket{0}, \ket{1}\} = 
\text{Span} \{\ket{b}, \ket{d}\}$ and its dynamics. The subspace evolves into $M(t)$, spanned by:
\begin{eqnarray}
\ket{\psi_k(t)} =\exp\left( - i \int_0^t H_I^{\rm RWA}(t') dt' \right) \ket{k}
= \mathcal{U}(t, 0) \ket{k},
\end{eqnarray}
where $\mathcal{U}(t, 0)$ is the time-evolution operator in RWA and $k=0$ and $1$. In the bright-dark basis, 
the unitary matrix assumes the form \cite{Herterich2016} 
\begin{eqnarray}
\mathcal{U} (t, 0)
& = &
\ket{d}\bra{d}+\cos \Phi (\ket{b}\bra{b} + \ket{e}\bra{e}) \nonumber \\ 
& & - i \sin \Phi (\ket{e}\bra{b} + \ket{b}\bra{e}),
\end{eqnarray}
where 
\begin{eqnarray}\label{eq:pulse_area_int}
\Phi
=
\int_0^t \Omega(t')dt'
\end{eqnarray}
is the pulse area.  Geometrically, this evolution corresponds to a path in the Grassmanian 
$\mathcal{G}(3;2)$, i.e., the set of two-dimensional subspaces of the three-dimensional 
Hilbert space that define the $\Lambda$-type system \cite{Bengtsson2006}. In particular, 
when we have $\Phi = \pi$, the trajectory corresponds to a full loop in the Grassmanian. 
The effect of this evolution whenever $\Phi = \pi$ is to implement a holonomy matrix 
\begin{eqnarray}\label{eq:single_pulse_gate}
U(C) = \mathcal{U} (C) \mathbb{P} = 
\begin{matrix}
\begin{pmatrix}
\cos \theta & e^{-i \phi } \sin \theta\\
 e^{i \phi } \sin \theta\ & -\cos \theta
\end{pmatrix}
\end{matrix}
=
{\bf n} \cdot \boldsymbol{\sigma} , 
\end{eqnarray}
after properly projecting $\mathcal{U}(t, 0)$ into the qubit space. Here, 
${\bf n} = (\sin \theta \cos \phi, \sin \theta \sin \phi, \cos \theta)$ and 
$\mathbb{P} = \ket{0}\bra{0} + \ket{1} \bra{1}$. Besides, we have  parametrized the 
frequencies $\omega_0$ and $\omega_1$ as $\omega_0=\sin(\theta/2)e^{i\phi}$ and 
$\omega_1=-\cos(\theta/2)$, i.e., $\theta$ and $\phi$ determine the relative amplitude and phase, respectively, of the two laser pulses. This process corresponds to a $\pi$ rotation around ${\bf n}$ 
on the Bloch sphere. The unitary in Eq.~(\ref{eq:single_pulse_gate}), however, is not 
universal (the phase-shift gate, for instance, cannot be written in this form). A universal 
gate can be implemented by employing a second 
loop $C_{{\bf m}}$, yielding 
\begin{eqnarray}\label{eq:ideal_unitary}
U(C) = U(C_{{\bf m}})U(C_{{\bf n}}) = {\bf n} \cdot {\bf m}
- i \boldsymbol{\sigma} \cdot ({\bf n} \times {\bf m}).
\end{eqnarray}
This transformation has a clear physical meaning as well; the universal gate $U(C)$ above 
corresponds to a rotation in the plane spanned by ${\bf n}$ and ${\bf m}$ by an angle of 
$2\cos^{-1}({\bf n} \cdot {\bf m})$ \cite{Sjoqvist2012}. Therefore, any single-qubit gate can 
be obtained by an appropriate choice of pulses, determined by ${\bf n}$ and ${\bf m}$. The 
holonomic nature of $U(C)$ in Eq.~\eqref{eq:ideal_unitary} relies on two facts \cite{Sjoqvist2012}: 
(i) the Hamiltonian matrix $\bra{\psi_k (t)} H_I (t) \ket{\psi_l (t)}$ vanishes so that $U(C)$ 
becomes purely dependent on $C$, and 
(ii) there exist two generic loops $C$ and $C'$ in $\mathcal{G} (3;2)$ for which the 
corresponding unitaries do not commute. The latter, together with the two-qubit setting 
described below, ensures universality of NHQC. 

From hereafter, we use hyperbolic secant pulses 
\begin{eqnarray}\label{eq:pulse_shape}
\Omega(t) = \beta \sech (\beta t),     
\end{eqnarray}
as depicted in Fig.~\ref{fig:setup_dissipative}(b). The 
parameter $\beta$ can be regarded as the inverse pulse length: by increasing $\beta$ we 
implement sharper and shorter pulses. It also provides a convenient parametrization: this 
pulse is area-preserving and the condition $\Phi = \pi$ is satisfied regardless of the value 
of $\beta$ chosen.

Finally, we use the amplitude-damping jump operator $L = \ket{g}\bra{e}$, with $\ket{g}$ being an 
additional low lying level, and the Lindblad equation 
\begin{eqnarray}\label{eq:RWAdissipative}
\frac{d\rho}{dt} = i[\rho, H_I(t)] + \gamma D(\rho), 
\end{eqnarray}
to model decay \cite{Breuer2007}. Here, $\rho$ is the density matrix of the system, 
$D(\rho)=L \rho L^\dagger -\frac{1}{2}\{L^\dagger L, \rho\}$ is the dissipator, and 
$\gamma$ is the dissipation rate, as depicted in Fig.~\ref{fig:setup_dissipative}(a).      

\subsection{Two-qubit gate} 

The original NHQC proposal in Ref.~\cite{Sjoqvist2012} includes a protocol based on 
the S\o rensen–M\o lmer scheme \cite{Sorensen1999} for implementing two-qubit gates 
(see also Ref.~\cite{Duan2001} for an adiabatic implementation and 
Refs.~\cite{ZhaoXu2019,Xu2021} for generalizations of the NHQC scheme to multi-qubit  
gates). The setup for the two-qubit gate consists of two identical trapped ions in the same 
three-level $\Lambda$-configuration as the one used for single-qubit gates. The transition 
$0 \leftrightarrow e$ ($1 \leftrightarrow e$) is driven by a laser with detuning 
$\pm (\nu + \delta)$ [$\pm (\nu - \delta)]$, where $\nu$ is the trap frequency and $\delta$ 
is an additional detuning, as shown in Fig.~\ref{fig:setup_dissipative}(c). In addition, we 
assume that the system satisfies the Lamb-Dicke criterion $\eta \ll 1$, where $\eta$ is the 
Lamb-Dicke parameter, and also that $|\Omega_i(t)| < \nu$. The parameter $\eta$ depends on the zero-point spread of the ion \cite{Wineland1998} and accounts for the coupling strength between its internal degrees of freedom and its motional states \cite{Li2002}. These conditions allow us to 
suppress the off-resonant couplings \cite{Sorensen1999}. As such, the Hamiltonian 
describing this interaction assumes the form
\begin{eqnarray}\label{eq:twoqubit_Hamiltonian}
    H^{(2)} (t) & = & 
    \frac{\eta^2}{\delta}\bigg(
    |\Omega_0(t)|^2 \sigma_0(\phi, t) \otimes \sigma_0(\phi, t) \nonumber \\
    & & - |\Omega_1(t)|^2 \sigma_1(-\phi, t) \otimes \sigma_1(-\phi, t)\bigg),
\end{eqnarray}
where
\begin{eqnarray}\label{eq:twoqubit_counter}
    \sigma_0(\phi, t) = e^{i \phi/4}(1 + e^{-2i f_{0e} t})\ket{e}\bra{0} + {\rm H.c.}, \nonumber \\
    \sigma_1(-\phi, t) = e^{-i \phi/4}(1 + e^{-2i f_{1e} t})\ket{e}\bra{1} + {\rm H.c}.
\end{eqnarray}
After eliminating off-resonant couplings of the singly excited states $\ket{0e}$ , $\ket{e0}$, 
$\ket{1e}$ and $\ket{e1}$, and once again performing the RWA, the Hamiltonian reads:
\begin{eqnarray}\label{eq:twoqubit_total}
    H^{(2) , {\rm RWA}}(t)
    =
    \sqrt{|\Omega_0(t)|^4 + |\Omega_1(t)|^4}
    \left( \tilde{H}^{(2)}_0 +
    \tilde{H}^{(2)}_1 \right), 
\end{eqnarray}
with
\begin{eqnarray}\label{eq:twoqubit1}
    \tilde{H}^{(2)}_0 
    =
    \sin{\frac{\theta}{2}} e^{i \phi/2} \ket{ee}\bra{00} 
    - \cos{\frac{\theta}{2}} e^{-i\phi/2} \ket{ee}\bra{11} + {\rm H.c.} \nonumber \\  
\end{eqnarray}
and
\begin{eqnarray}\label{eq:twoqubit2}
    \tilde{H}^{(2)}_1 
    =
    \sin{\frac{\theta}{2}} \ket{e0}\bra{0e} - \cos{\frac{\theta}{2}}  \ket{e1}\bra{1e} +  {\rm H.c.}  
\end{eqnarray}
Here, $|\Omega_0(t)|^2/|\Omega_1(t)|^2 = \tan(\theta/2)$ and $\phi$ are kept constant 
throughout the pulse. Likewise, we should assume the criterion 
$\frac{\eta^2}{\delta} \int_0^\tau \sqrt{|\Omega_0(t)|^4 + |\Omega_1(t)|^4}dt = \pi$ for the 
pulse area is satisfied. By analogy to the single-qubit gate, this procedure yields the unitary
\begin{eqnarray}
    U^{(2)}(C_n) & = &
    \cos{\theta}\ket{00}\bra{00} + e^{-i\phi}\sin{\theta}\ket{00}\bra{11} \nonumber\\
    & & + e^{i\phi}\sin{\theta}\ket{11}\bra{00} - \cos{\theta}\ket{11}\bra{11} \nonumber\\
    & & + \ket{01}\bra{10} + \ket{10}\bra{10}.
\end{eqnarray}
By choosing $\theta = 0$, we are able to construct a controlled-$Z$ (CZ) gate
\begin{eqnarray}
    U^{(2)}_{\rm CZ}
    =
    \ket{00}\bra{00}
    +\ket{01}\bra{10}
    +\ket{10}\bra{10}
    -\ket{11}\bra{11},
\end{eqnarray}
which is an entangling gate and can form a universal set together with a universal 
single-qubit gate \cite{Bremner2002}.
 
\begin{figure}
     \center
     \includegraphics[width=\columnwidth]{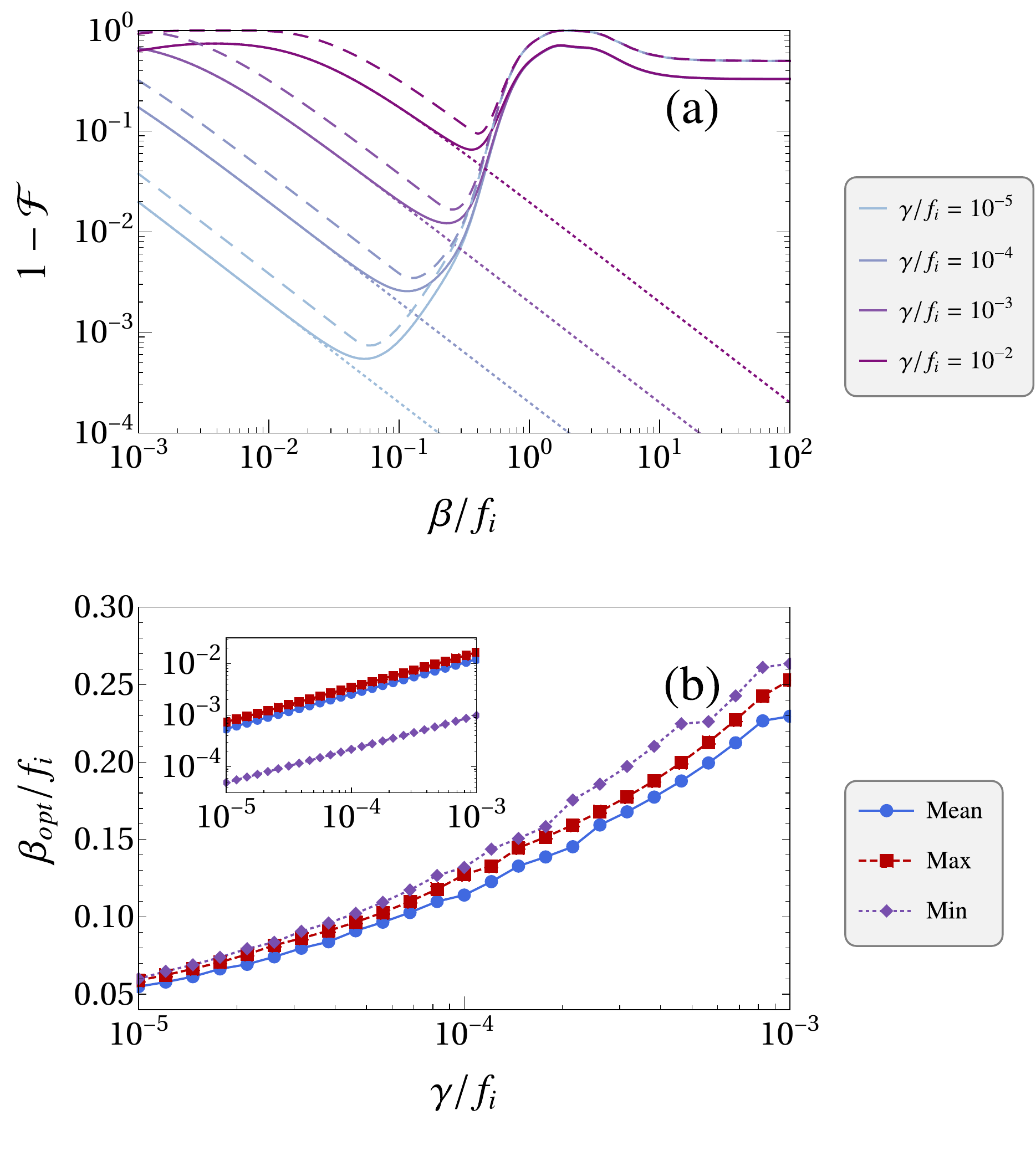}
     \caption{
     (a) The solid (dashed) lines correspond to the mean (maximum) infidelity $1 - \mathcal{F}$ of the $S$ gate as a function of the ratio $\beta/f_i$ for 
     different values of $\gamma/f_i$. The dotted lines correspond to the RWA solution 
     where the Hamiltonian in Eq.~\eqref{eq:Lambda_hamiltonian_rwa} is used together 
     with the master equation in Eq.~\eqref{eq:RWAdissipative}, i.e., when counter-rotating 
     effects are neglected. The spacing between the two pair of pulses is chosen as 
     $\Delta t = 10/\beta$. This guarantees that the pulses are not significantly overlapping. 
     We used a grid of 500 sample points for $\beta/f_i$. (b) Optimal inverse pulse length 
     $\beta_{opt}$ (in units of $f_i$) as a function of $\gamma/f_i$. We also plot the 
     corresponding infidelity (inset).  We have used a sample space of 1000 points for 
     $\beta_i/f_i$ in the interval $[0.03, 0.3]$. The average infidelity was calculated for 
     $100$ input states uniformly sampled over the Bloch sphere in both figures.}
     \label{fig:FidelityPhaseShift}
\end{figure}

\section{Results}\label{sec:nonRWA}
 
 \begin{figure*}[htp!]
     \center
     \includegraphics[width=0.98\textwidth]{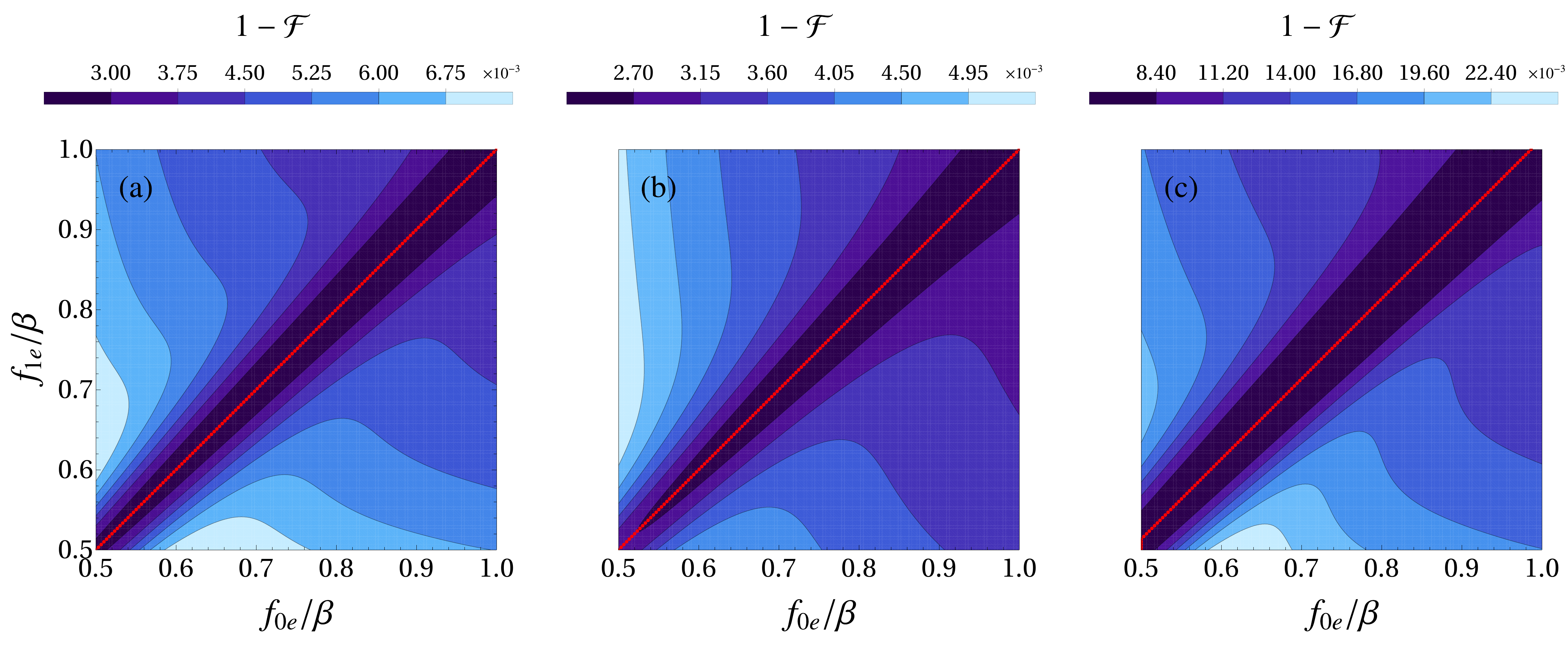}
     \caption{
    Average infidelity $1-\mathcal{F}$ as a function of the counter-rotating frequencies 
    $f_{0e}$ and $f_{1e}$ for $\gamma/\beta = 0.02$. Results are shown for (a) the $X$ gate, 
    (b) the Hadamard gate, and (c) the $S$ gate. Other details and parameters are the same 
    as those in Fig.~\ref{fig:FidelityPhaseShift}. The simulations were done for a grid of $150 \times 150$ 
    frequencies $f_{0e}$ and $f_{1e}$.}
     \label{fig:FidelityGrid}
\end{figure*}

The validity of the RWA has been studied previously in the dissipationless case for 
single-qubit gates  \cite{Spiegelberg2013}. There, the authors show that the approximation 
starts to break down for very short pulses. We further extend this investigation, considering 
the competing effect of dissipation and counter-rotating terms; while faster gates avoid 
decoherent effects, we also observe the counter-rotating terms playing a larger role in 
the dynamics, compromising the fidelity of the gate.  The converse is also true: if we use 
longer duration pulses in order to offset the counter-rotating terms, the accuracy suffers 
due to the longer exposure time to dissipative effects. Our objective is to find an optimal 
relationship between these parameters, namely, to investigate the interplay among three 
components in this model: the inverse pulse length $\beta$ defining the hyperbolic secant pulse in Eq.~(\ref{eq:pulse_shape}), the coupling parameter $\gamma$, 
and the counter-rotating frequencies $f_{je}$.

If we write Eq.~\eqref{eq:Lambda_hamiltonian} in terms of the bright and the dark states we obtain 
\begin{eqnarray}\label{eq:Hbd_rotating}
    H_{I}(t) & = &
    \Omega(t)
    (1+|\omega_0|^2e^{-2i f_{0e} t} + |\omega_1|^2e^{-2i f_{1e} t} )
    \ket{e}\bra{b} \nonumber \\
    & & \quad + \Omega (t)\omega_0 \omega_1(e^{-2i f_{1e} t} - e^{-2i f_{0e} t})
\ket{e}\bra{d}
+\mathrm{H.c.}
\end{eqnarray}
Differently from what we observe in the ideal case, the presence of counter-rotating terms 
introduces a coupling between the dark and the excited state. An exception occurs in the case 
of homogeneous frequencies, that is, when $f_{1e}=f_{0e}=f_i$, resulting in 
\begin{eqnarray}\label{eq:Hbd_rotating2}
    H_{I}(t) =
    \Omega(t)(1+e^{-2i f_i t} )\ket{e}\bra{b}
    +\mathrm{H.c.}
\end{eqnarray}
The equation above means that taking the two counter-rotating frequencies to be the same 
eliminates the coupling between the dark state and the excited state in Eq.~\eqref{eq:Hbd_rotating}. 
On the other hand, we can see that Eq.~\eqref{eq:Hbd_rotating2} differs from
Eq.~\eqref{eq:Lambda_hamiltonian_rwa_bd} by a factor of $e^{-2i f_i t}$ in the Rabi 
frequency. This means that even though the structure of the Hamiltonian is the same 
as the one found in the ideal case, due to the counter-rotating corrections we do not, in 
general, return to the initial subspace at the end of the evolution.  

\subsection{Singe-qubit gates}\label{sec:optimal_configuration}

In order to investigate the validity of the RWA, we compute the fidelity $\mathcal{F} = \bra{\psi_0}
U(C)^\dagger \rho U(C) \ket{\psi_0}$ for different quantum gates, and analyze the simulation 
results in terms of the infidelity $1-\mathcal{F}$. The unitary $U(C)$ corresponds to the ideal 
unitary in Eq.~\eqref{eq:ideal_unitary}, which is determined by the choice of ${\bf n}$ and 
${\bf m}$. The density matrix $\rho$ is obtained by solving the master equation,   
Eq.~\eqref{eq:RWAdissipative}, for the non-RWA Hamiltonian in 
Eq.~\eqref{eq:Lambda_hamiltonian}, and  $\ket{\psi_0}$ is the input state. Finally, we 
compute the \emph{average} fidelity for several input states uniformly distributed over 
the Bloch sphere. For details, see the Appendix.

We start by examining the average infidelity for the phase-shift gate. By choosing 
$\theta = \theta' = \pi/2$ resulting in the pulses ${\bf n} = (\cos \phi,\sin \phi,0)$ and 
${\bf m} = (\cos \phi', \sin \phi', 0)$, we can implement the phase-shift gate 
$\ket{k} = e^{2ik(\phi' - \phi)} \ket{k}$, with $k=0$ and $1$. Basic results are shown in  
Fig.~\ref{fig:FidelityPhaseShift}(a) for the average infidelity $1 - \mathcal{F}$ as a 
function of the ratio between the inverse pulse length $\beta$ and the counter-rotating 
frequencies taken to be the same, i.e., $f_{0e} = f_{1e} = f_i$. In Fig.~\ref{fig:FidelityPhaseShift}(b), we also include the maximum and minimum infidelities in order to investigate how the gate might behave depending on the input state. They seem to play a minor role regarding the optimal configuration of the gate. While we observe a significant gap in the performance of the maximum and minimum achievable infidelities in the inset of Fig.~\ref{fig:FidelityPhaseShift}(b), the infidelity as a function of $\beta/f_i$ behaves very similarly for different input states and the optimal inverse pulse length in Fig.~\ref{fig:FidelityPhaseShift}(b) remains largely unchanged, specially at lower frequencies. Therefore, the mean infidelity is, by itself, able to capture the general behavior of the gates and its optimal configurations sufficiently well.

We can notice that for 
small $\beta/f_i$ (longer pulses) the curves should converge roughly to the same fidelity, but 
at different rates depending on $\gamma/f_i$. This is not visible in the plot for smaller rates of $\gamma/f_i$ due to the increasingly large timescales necessary to observe this effect, which make numerical simulations difficult.  This happens due to the fact that in this 
regime the decoherent effects become dominant. More specifically, the dissipation 
hampers any population transfer between the qubit states, as in this case the 
populations and coherences in the excited state quickly decay to the ground state. 
Therefore, the final state is very nearly  the same as the input state, regardless of 
the value of $\beta/f_i$. We make a quantitative support to this claim at the end of this section.

On the other hand, for large $\beta/f_i$ (shorter pulses) we can see that the gate 
accuracy also decreases and the fidelity becomes independent of the ratio $\gamma/f_i$. 
In this scenario the operation time is very short, hence the counter-rotating terms dominate 
and the end result depends on the particular value of $f_i$. The asymptotic behavior 
for $\gamma \ll f_i \ll \beta$ has been briefly discussed in Ref.\cite{Spiegelberg2013} for the $X$ 
and the Hadamard (${\rm H}$) gates, when considering a few selected input states. The 
counter-rotating correction in Eq.~\eqref{eq:Hbd_rotating2} becomes $1 + e^{-2 i f_{i} t} 
\approx 2$ for small $f_{i}$. In this regime,  the system undergoes a cyclic evolution twice. 
This means that the bright state evolves as $\ket{b} \rightarrow -\ket{b} \rightarrow \ket{b}$, 
and the net effect of the evolution is simply to (approximately) implement the identity gate, 
once again leaving the input state unchanged. 

Finally, for values in-between we observe the optimal regime for the quantum gate and 
the presence of a global minimum in the infidelity $1-\mathcal{F}$. This is consistent 
with the intuitive idea behind the competition between the RWA and the dissipative effects: 
the optimal pulse duration should be long enough in order to minimize the counter-rotating 
contributions to the dynamics while also being short enough to avoid dissipative losses. 
After that point the RWA starts to break down and the infidelity increases. Figure 
\ref{fig:FidelityPhaseShift}~(a) explicitly shows an overlap between the RWA solution 
(dashed line) and the full solution (solid lines). The latter deviates from the former when 
$\beta/f_i$ approaches the global minimum, explicitly showing a breakdown of the RWA. 
Therefore, for optimal performance, the frequencies and the pulse should be tuned in a 
way that $f_i > \beta \gg \gamma$ holds, if possible. This means that the inverse pulse 
length $\beta$ should be much larger than $\gamma$, and at the same time the 
counter-rotating frequencies should be sufficiently larger than $\beta$. 

\subsection{Heterogeneous frequencies}\label{sec:het_freq}

We also examine how different counter-rotating frequencies affect the fidelity of the gates. 
In Fig.~\ref{fig:FidelityGrid} we show a contour plot for the infidelity $1-\mathcal{F}$ as a 
function of $f_{je}/\beta$ for three different gates. Results are shown for the $X$, 
${\rm H}$, and $S$ gates, respectively. The red lines correspond to the optimal 
frequency $f_{1e}$ for a given $f_{0e}$. A few observations can be made from these plots. 
Foremost, we can explicitly see that for the range of frequencies considered here the $S$ 
gate has a worse fidelity in general. This corroborates the fact that, since the $S$ gate 
requires two pulses, the longer operation time makes the gate much more susceptible 
to open quantum system effects. Additionally, it is also possible to observe that 
Fig.~\ref{fig:FidelityGrid}(a) is symmetric, while Fig.~\ref{fig:FidelityGrid}(b) is not. This is linked to the fact 
that the complex frequencies $\omega_0$ and $\omega_1$ are of the same modulus 
for the $X$ gate, but not for the Hadamard gate. 

Furthermore, we can analyze the optimal combinations of frequencies in 
Fig.~\ref{fig:FidelityGrid}. The results shown in Figs.~\ref{fig:FidelityGrid}(a) and ~\ref{fig:FidelityGrid}(b) show that for single pulse gates and for the frequency range considered, the 
optimal relationship for the counter-rotating frequencies is to take both of them to 
be the same, i.e., the infidelity $1-\mathcal{F}$ is minimum for $f_{0e} = f_{1e}$. 
This is possibly a result of the decoupling of the dark and the excited states, which 
happens in Eqs.~\eqref{eq:Hbd_rotating} and \eqref{eq:Hbd_rotating2}. The coupling 
between $\ket{e}$ and $\ket{d}$ in Eq.~\eqref{eq:Hbd_rotating} seems to play a larger 
role than the presence of the counter-rotating correction in Eq.~\eqref{eq:Hbd_rotating2}. 
Increasing one of the frequencies while keeping the other to be the same shows no 
noticeable improvement. On the contrary, we observe a slight worsening of the gate 
performance. 

Surprisingly, we observe a different behavior for the $S$ gate: there is a slight offset 
between the optimal $f_{0e}$ and $f_{1e}$. The result indicates that ideally one should 
increase these two frequencies at a constant ratio,  different from unity. This behavior 
may arise from the use of two pair of pulses in the $S$ gate implementation. It has been 
verified in Ref.~\cite{Spiegelberg2013} that the fidelity of non-commuting gates is actually 
lower than the product of their fidelities. A similar mechanism may be playing a role here: 
the non-Abelian behavior of the unitaries in Eq.~\eqref{eq:ideal_unitary} is possibly the 
reason why we observe this effect for unequal frequencies in the S gate. 

\begin{figure}[t!]
     \center
     \includegraphics[width=0.95\columnwidth]{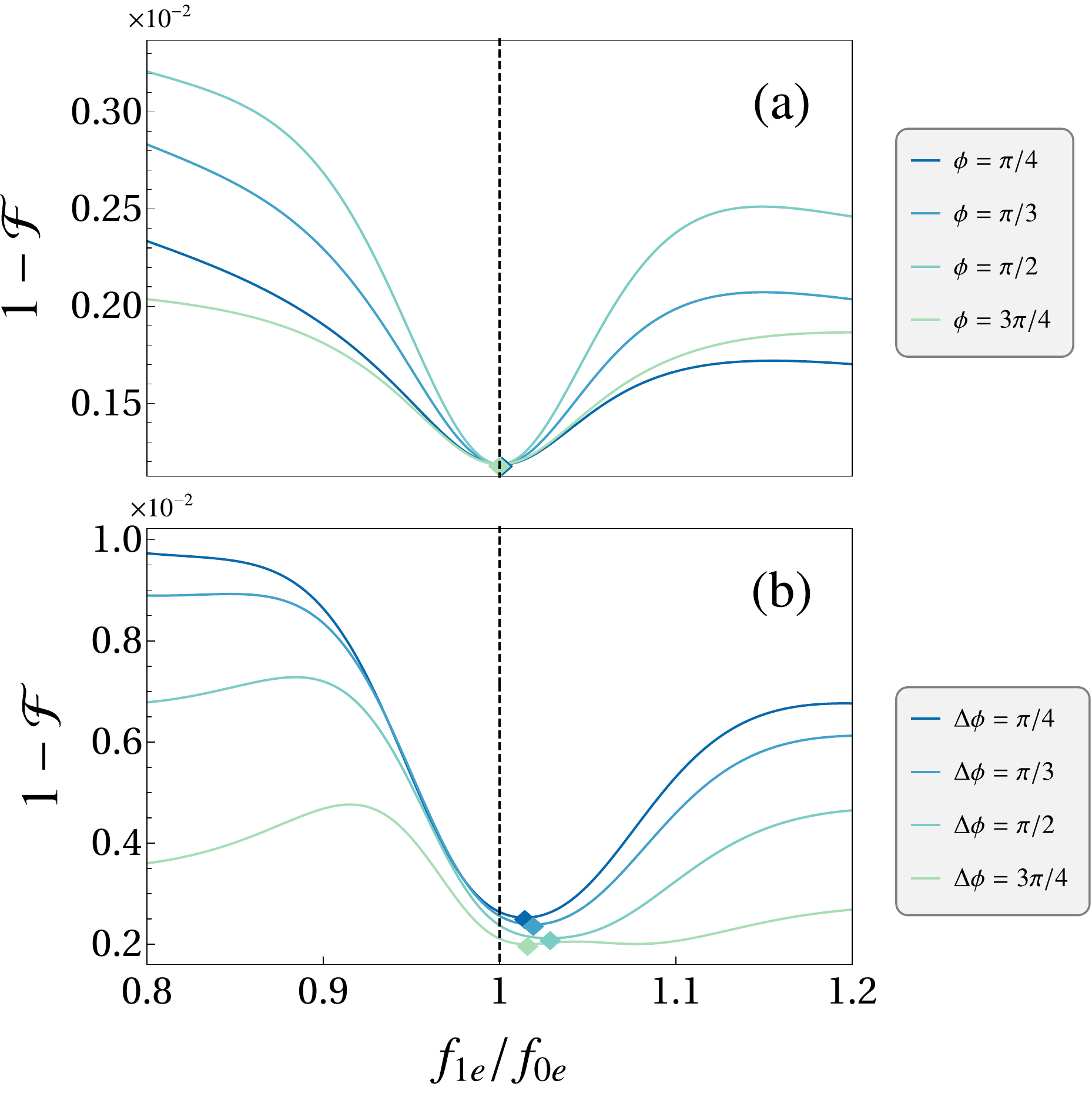}
     \caption{
     Infidelity as a function of the ratio $f_{1e}/f_{0e}$ for single-qubit gates. We show the 
     results for (a) a single-pulse gate with $\theta = \pi/4$, for different values of $\phi$. 
     In panel (b) we plot the results for the phase-shift gate, for which $\theta = \theta' = \pi/2$, 
     for different values of $\Delta \phi \equiv \phi' - \phi$. We observe a slight shift in the 
     optimal ratio for the phase-shift gates. The optimal points depends on the desired phase. 
     Here, the coupling strength is given by $\gamma/\beta = 10^{-3}$, the spacing between 
     the pulses is $\Delta t = 20/\beta$ and the frequency $f_{0e}$ is fixed and given by 
     $f_{0e}/\beta = 10$. We use a sampling space of $250$ frequencies and $100$ input 
     states, as described in Fig.~\ref{fig:FidelityPhaseShift}.} 
     \label{fig:OptimalFreqPlot}
\end{figure}

\begin{figure}[t!]
     \center
     \includegraphics[width=\columnwidth]{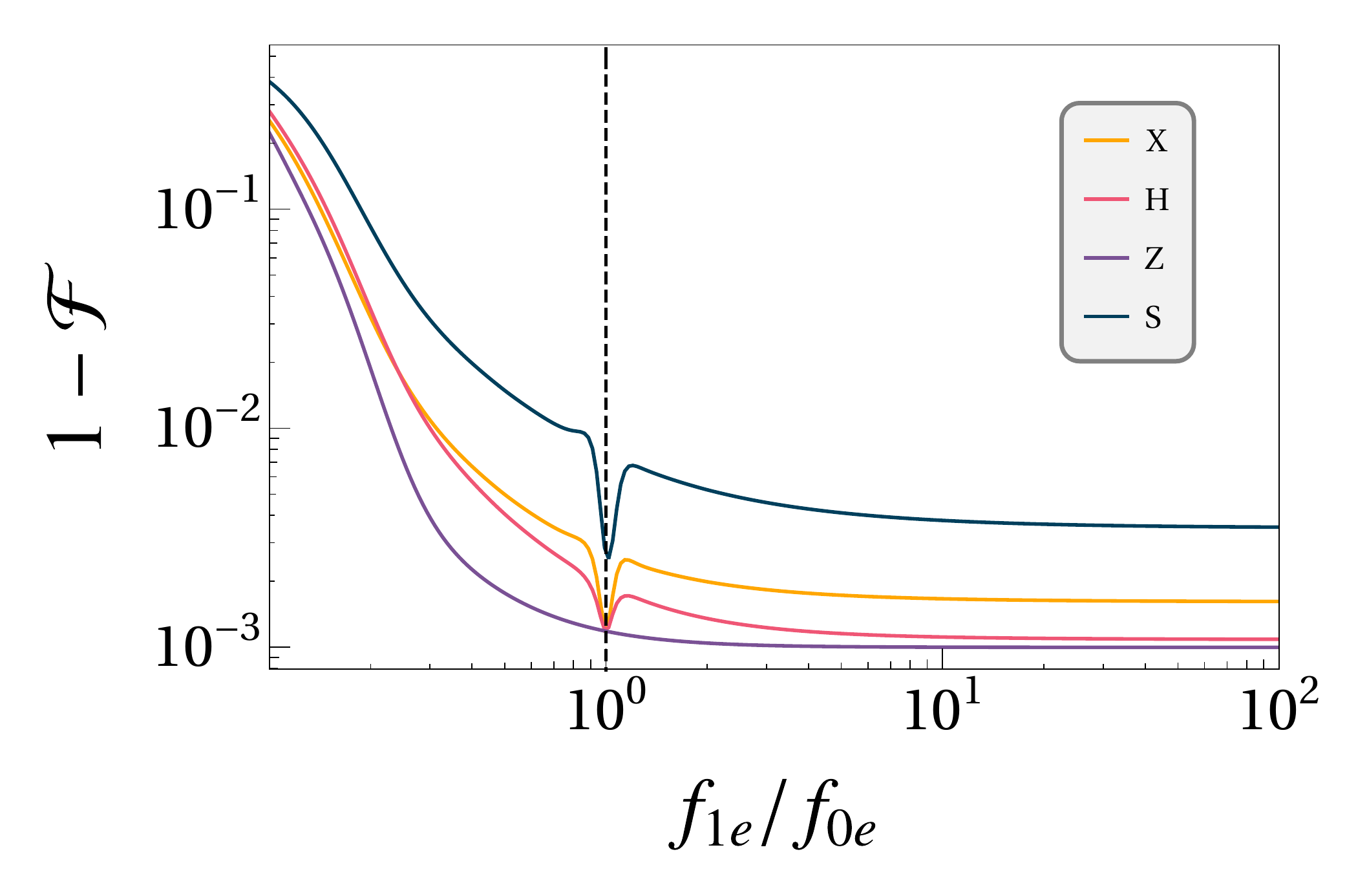}
     \caption{
     Infidelity as a function of the ratio $f_{1e}/f_{0e}$ for the {\rmfamily \scshape  NOT}  gate (yellow), the Hadamard 
     gate (magenta), the $Z$ gate (purple) and the $S$ gate (dark blue). The decay rate, the sampling 
     space and $f_{e0}$ are the same as those in Fig.~\ref{fig:OptimalFreqPlot}. Note that the minimum 
     for the $S$ gate is slightly displaced from $f_{1e}/f_{0e} = 1$ (dashed line), in accordance 
     with Figs.~\ref{fig:FidelityGrid}(c) and~\ref{fig:OptimalFreqPlot}.}
     \label{fig:FrequencyLog}
\end{figure}

Now, we further investigate the effect of heterogeneous frequencies. For that we consider 
the infidelity of two different single qubit gates. For the first gate we implement the single-pulse 
gate in Eq.~\eqref{eq:single_pulse_gate} for different values of $\phi$, with $\theta = \pi/4$. 
For the second gate, we implement the phase-shift gate for different phases 
$\Delta \phi = 2(\phi' - \phi)$. Our objective is to analyze how changing $\phi$ affects the 
optimal ratio between the two counter-rotating frequencies. Like Fig.~\ref{fig:FidelityGrid}, 
the curves in Fig.~\ref{fig:OptimalFreqPlot}(a) show that single-pulse gates achieve optimality 
for equal frequencies. On the other hand, Fig.~\ref{fig:OptimalFreqPlot}(b) shows that while 
the optimal ratio indeed occurs for different frequencies, the gain is very marginal. This 
suggests that a weak coupling between the excited state and the dark state alone is much 
more relevant overall. This claim is further supported by the result in Fig.~\ref{fig:FrequencyLog}. 

We see, for the $X$, ${\rm H}$ and $S$ gates, that there is a sudden decrease 
in the infidelity when $f_{0e} \approx f_{1e}$. For the $S$ and the $X$ gates 
the point $f_{0e} = f_{1e}$ is actually a global minimum. Meanwhile, when the ${\rm H}$ 
gate is considered, the point where the two frequencies are the same only happens to be 
a \emph{local} minimum. For the ${\rm Z}$ gate we observe no minimum at all, and the 
infidelity is simply monotonically decreasing with $f_{1e}/f_{0e}$. Hence, whether the point 
$f_{0e} = f_{1e}$ is optimal may depend on the particular gate implementation and on the 
frequency range considered. Nevertheless, a characteristic drop in the infidelity around 
this region seems to be common among all of the gates but the $Z$ gate, and the 
performance in the regime where $f_{0e} = f_{1e}$ is quite similar to the one obtained 
when $f_{1e} \gg f_{0e}$. The behavior of the curves in Fig.~\ref{fig:FrequencyLog} 
arises mainly because of the last term in Eq.~\eqref{eq:Hbd_rotating}. For the ${\rm Z}$ 
gate, for instance, we have $\omega_0 = 0$ and $\omega_1 = 1$. In this very particular 
case, this term vanishes and the fidelity actually depends only on $f_{1e}$, and we do 
not observe any coupling between the dark state and the excited state regardless of $f_{0e}$. 
Meanwhile, the $X$ gate is quite different: since $|\omega_0| = |\omega_1| = 1/\sqrt{2}$, 
the dependence of the infidelity on $f_{0e}$ and $f_{1e}$ is symmetric and the product 
$\omega_0 \omega_1$ is maximum (in modulus), thus, the $X$ gate (or any other 
gate for which $|\omega_0| = |\omega_1|$, for that matter) is the most sensitive to the 
coupling between the dark state and the excited state. For other gates, such as the Hadamard 
gate, we observe a behavior in-between. The ratio $|\omega_1|/|\omega_0|$ between 
the amplitudes should then play some role whether the optimal strategy is to take 
$f_{1e}/f_{0e} = 1$ or $f_{1e} \gg f_{0e}$ for maximum fidelity.

\subsection{Two-qubit gates}\label{sec:two_qubit}

\begin{figure}
     \center
     \includegraphics[width=\columnwidth]{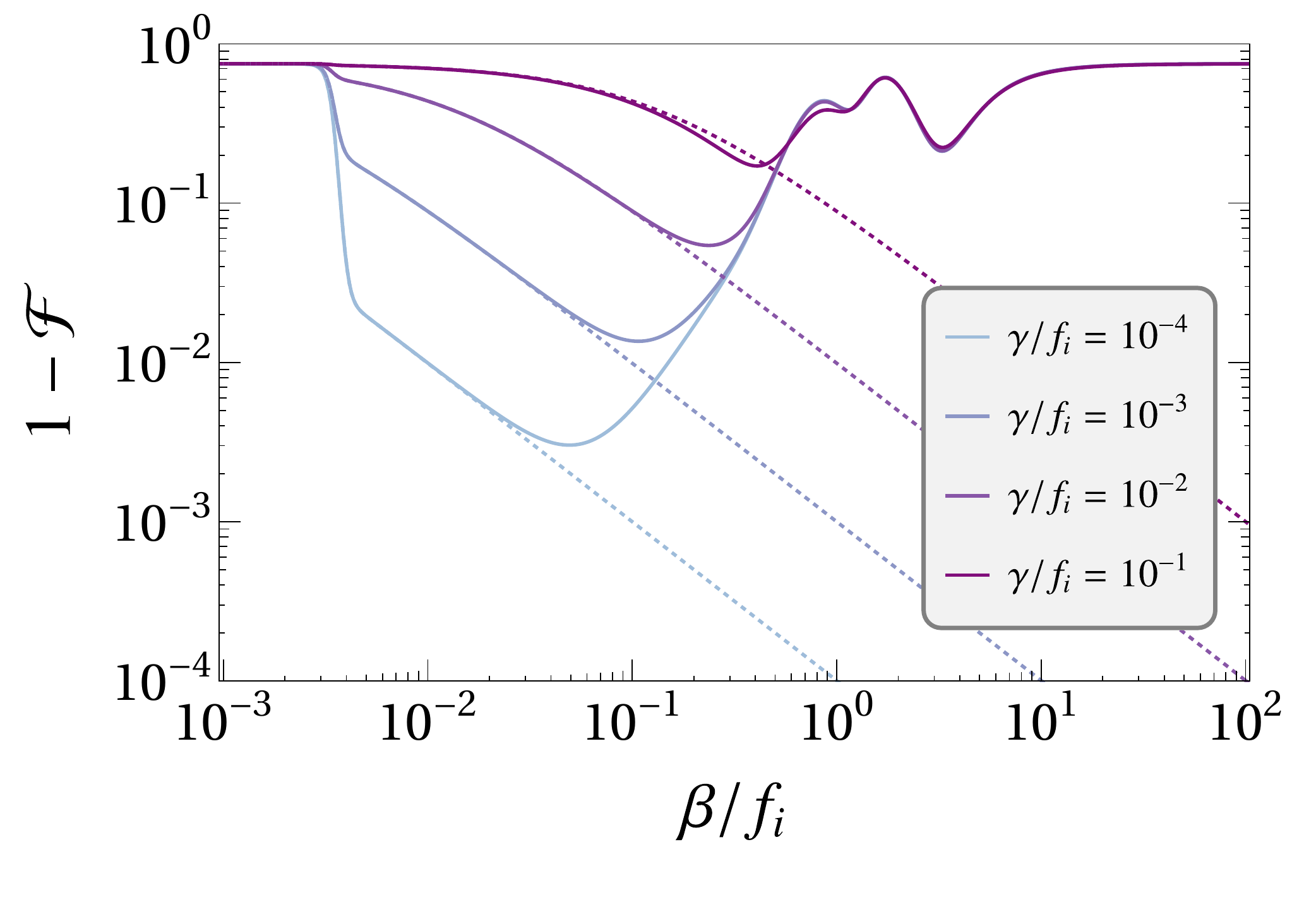}
     \caption{
     Infidelity of the CZ gate as a function of $\beta/f_i$. Dotted lines represent the RWA 
     solution given by the Hamiltonian in Eq.~\eqref{eq:twoqubit1}. Other details are the 
     same as those in Fig.~\ref{fig:FidelityPhaseShift}.}
     \label{fig:CZFidelity}
\end{figure}

Finally, we check the robustness against both effects in the case of the two-qubit gate. 
Considering the counter-rotating contributions in Eqs.~\eqref{eq:twoqubit_Hamiltonian} 
and \eqref{eq:twoqubit_counter}, Eqs.~\eqref{eq:twoqubit1} and \eqref{eq:twoqubit2} become
\begin{eqnarray}
    H^{(2)}_0(t) & = &
    (1 +  e^{-2 i f_{0e} t})^2\sin{\frac{\theta}{2}} e^{i \phi/2} \ket{ee}\bra{00} \nonumber \\
    & & - (1 +  e^{-2 i f_{1e} t})^2 \cos{\frac{\theta}{2}} e^{-i\phi/2} \ket{ee}\bra{11} + {\rm H.c.}
\end{eqnarray}
and
\begin{eqnarray}
    H^{(2)}_1(t) & = &
    4 \cos^2{(f_{0e} t)} \sin{\frac{\theta}{2}} \ket{e0}\bra{0e} \nonumber \\ 
    & & - 4 \cos^2{(f_{1e} t)} \cos{\frac{\theta}{2}}  \ket{e1}\bra{1e} + {\rm H.c.},
\end{eqnarray}
respectively. We perform simulations for the CZ gate, which can be constructed by taking 
$\theta = 0$. Results for the infidelity are shown in Fig.~\ref{fig:CZFidelity} as 
a function of the inverse pulse length $\beta$ (in units of $f_{1e} = f_{0e} = f_i$). We plot 
the infidelity, averaged over four different input states: $\ket{+}\ket{+}$, $\ket{+}\ket{-}$, 
$\ket{-}\ket{+}$ and  $\ket{-}\ket{-}$ with $\ket{\pm}  = (\ket{0} \pm \ket{1})/\sqrt{2}$, which 
are states of interest since the application of the CZ gate upon them results in maximally 
entangled Bell states. Our results are qualitatively similar to what was obtained for the single-qubit 
case, showing that the two-qubit gate is also robust against the joint effect of dissipation and 
counter-rotating terms. Moreover, the asymptotic behavior for larger pulses is even more 
evident in Fig.~\ref{fig:CZFidelity}; we can clearly see how the dissipative effects quickly 
start to dominate for $\beta/f_i \lesssim 4 \cdot 10^{-3}$, eventually converging to a specific 
value, in a similar fashion to what we obtained for the phase-shift gate. The same 
phenomenon occurs when $\beta \gtrsim f_i$ due to the increasing influence of counter-rotating 
contributions. This behavior arises precisely due to the two mechanisms we have 
discussed before: for both the very fast or the very slow gate operation regimes, a given 
input state $\ket{\psi_0}$ is left roughly unchanged after the evolution, and the fidelity 
associated with the process is approximately $|\bra{\psi_0}U(C)\ket{\psi_0}|^2$. One 
may then find the asymptotic value for the average infidelity $1-\mathcal{F}$ in 
Figs.~\ref{fig:FidelityPhaseShift} and~\ref{fig:CZFidelity} by averaging 
$|\bra{\psi_0}U(C)\ket{\psi_0}|^2$ over all the input states. It is possible to show that 
$1 - \mathcal{F} = 1/3$ for the phase-shift gate in either cases. This is not visible in 
Fig.~\ref{fig:FidelityPhaseShift}~(a) simply due to the fact that we would need to use 
pulses of the order of $\beta/f_i \approx 10^{-4}$ or smaller in order to observe such a 
strongly dissipative regime. Moreover, by comparing Figs.~\ref{fig:FidelityPhaseShift} 
and~\ref{fig:CZFidelity} we could think that single and two-qubit gates are similarly 
robust, but that is not necessarily the case. As we can see in Figs.~\ref{fig:FidelityGrid}(a) 
and~\ref{fig:FidelityGrid}(b) and also in Fig.~\ref{fig:FrequencyLog}, single-pulse gates typically have lower 
infidelities. On the other hand, a positive feature is that the optimal pulse-length seems 
quite similar for both single- and two-qubit gates, which might simplify experimental 
realizations of holonomic quantum gates.

\subsection{Experimental feasibility}\label{sec:experimental}

On a closing note, we briefly discuss the experimental feasibility of an investigation 
regarding the frequency asymmetry. In Ref.~\cite{AbdumalikovJr2013}, for instance, the 
experiment is performed for a ratio of $f_{0e}/f_{1e} \approx 1.04$ between the frequencies 
and a dissipation rate given by $\gamma/f_i \approx 1.1 \cdot 10^{-4}$. As we can see in 
Fig.~\ref{fig:OptimalFreqPlot}, these parameters are close to the optimal scenario for both 
single and two-pulse gates. However, fine-tuning the frequencies to be the same in this 
setup in order to achieve even lower infidelities could be somewhat challenging. This 
implementation uses a transmon qubit based on a nearly harmonic ladder, where its 
ground state and second excited state span the qubit subspace and the first excited state 
plays the role of the auxiliary state $\ket{e}$. The anharmonicity is used to average out 
an undesirable coupling between the $\Lambda$-type system subspace and higher energy 
levels. Thus, homogeneous frequencies could possibly give rise to unwanted dynamical 
contributions. In this sense, implementations based on ion traps could be used to fine-tune 
the bare energy levels as desired, reaching the configuration shown in 
Figs.~\ref{fig:OptimalFreqPlot} and~\ref{fig:FrequencyLog} for the gates where 
homogeneous frequencies are optimal.

\section{Conclusions}\label{sec:conclusions}

We have investigated the robustness of non-adiabatic holomonic quantum gates against 
dissipative effects and counter-rotating dynamical contributions due to the energy structure 
of the $\Lambda$-type system. While short run times minimize the dissipative losses, the 
counter-rotating frequencies induce oscillations, which compromise the geometric character 
of the gate, decreasing its overall fidelity. In a real implementation, one should seek to 
sufficiently approach the RWA regime while keeping losses due to open quantum system 
effects minimal. In this work we have explored this trade-off in detail, obtaining an optimal 
regime of operation for single and two-qubit gates for a range of parameters typically found 
in experimental scenarios. Thus, we believe that our analysis could provide a suitable 
rule-of-thumb for optimal experimental parameters, as well as open up for optimization 
of gate performance by varying the pulse length, in implementations of quantum gates 
in the $\Lambda$-setting.

Moreover, we have also studied how different frequencies affect the performance of 
single-qubit gates. We showed that counter-rotating corrections introduce an unwanted 
coupling between the dark state and the excited state in the $\Lambda$-type system. Homogeneous 
frequencies in general result in a sudden decrease of the infidelity due to the suppression 
of the dark state coupling. However, whether this regime is indeed optimal will depend on 
the gate which is being implemented. In addition, we have observed that two-pulse 
single-qubit gates might display an optimal configuration for slightly heterogeneous 
frequencies. We suspect that this behavior arises due to the non-Abelian property 
of the geometrical phases.  This phenomenon can also be observed for square pulses 
in a slightly different fashion. However, whether our results still hold for arbitrary pulses 
shapes remain an open question and is subject for further investigation.

Future directions of study could either probe into generalizations of this analysis to 
different implementations of the holonomic protocol, such as the off-resonant scheme 
\cite{Xu2015,Sjoqvist2016}, the single loop scheme \cite{Herterich2016}, time-optimal-control   \cite{Chen2020,Han2020} and path shortening \cite{Xu2018b} schemes, and environment assisted implementations \cite{Ramberg2019}, or into inclusions of other open system 
effects, such as dephasing and correlated two-qubit noise.  A generalization to multi-qubit 
gates \cite{ZhaoXu2019,Xu2021} or strategies which do not rely on the RWA 
\cite{Gontijo2020} could also be investigated.


\section*{Acknowledgment}
G.O.A. acknowledges the financial support from the S\~ao
Paulo founding agency FAPESP (Grant No. 2020/16050-0)
and the framework of Erasmus+KA107 – International Credit Mobility, financed by the 
European Commission. E.S. acknowledges financial support from the Swedish Research 
Council (VR) through Grant No. 2017-03832.

\appendix
\setcounter{secnumdepth}{0} 
\section{APPENDIX: UNIFORM SAMPLING OVER THE BLOCH SPHERE}

Here, we show how the Fibonacci nodes algorithm \cite{Hardin2016} can be implemented. 
The algorithm can be used to uniformly sample states over the Bloch sphere. Canonical 
approaches for this problem based on, e.g., random sampling \cite{Muller1959} can also 
be used. However, the method presented here is deterministic, making it possible to get 
a reasonable approximation even for a smaller number of points. This greatly reduces the 
time required for computing the average fidelity, since we can perform simulations for a 
smaller number of input states. To implement the algorithm one should follow the steps below:
\begin{enumerate}
  \item Uniformly distribute the $z$-coordinate $z_n$ of the points in the interval $[1, -1]$.
  \item Distribute the azimuthal angles as $\phi_n = 2 \pi \phi n$, where $\phi = (1 + \sqrt{5})/2$ 
  is the golden ratio.
  \item Take $x_n = \sqrt{1 - z_n^2} \cos{\phi_n}$ and $y_n = \sqrt{1 - z_n^2} \sin{\phi_n}$.
  \item The points are given by the set of triples $(x_n, y_n, z_n)$, for $n = 1,...,N$.
\end{enumerate}
This set of triples can then be used to distribute the input states over the Bloch sphere.

\end{document}